\documentclass{aa}
\usepackage{graphicx,natbib}
\bibpunct{(}{)}{;}{a}{}{,}

\begin{document}

\sloppypar

%   \thesaurus{08     % A&A Section 6: Form. struct. and evolut. of stars
%              (02.01.2;  % Accretion, accretion disks
%               02.09.1;  % Instabilities
%              08.02.3;   % Stars:binaries:general
%              08.06.3;   % Stars:classification
%              08.14.1;   % Stars: neutron
%               13.25.3;  % X-rays: general
%               13.25.5)} % X-rays: stars
%

\title{Boundary layer emission and Z-track in the color-color
diagram of luminous LMXBs}

   \author{M.G. Revnivtsev\inst{1,2},
M.R. Gilfanov\inst{1,2},
}

   \offprints{mikej@mpa-garching.mpg.de}

   \institute{Max-Planck-Institute f\"ur Astrophysik,
              Karl-Schwarzschild-Str. 1, D-85740 Garching bei M\"unchen,
              Germany,
        \and
              Space Research Institute, Russian Academy of Sciences,
              Profsoyuznaya 84/32, 117997 Moscow, Russia
            }
  \date{}

        \authorrunning{Revnivtsev \& Gilfanov}
        \titlerunning{}

   \abstract{}
%Aims
{We explore the accretion disk and boundary layer emission in bright
neutron star LMXBs and their dependence on the mass accretion rate.
}
%methods
{Fourier-frequency resolved spectroscopy of the archival RXTE data.
}
%results
{
We demonstrate that Fourier-frequency resolved spectra
of atoll and Z- sources are identical, despite significant difference
in their average spectra and luminosity (by a factor of $\sim 10-20$).
This result fits in the picture we suggested earlier, namely that the
$f\ga 1$ Hz  variability in luminous LMXBs is primarily due to
variations of the boundary layer luminosity. In this picture the
frequency resolved spectrum equals the boundary layer spectrum, which
therefore can be straightforwardly determnined from the data.
The obtained so boundary layer spectrum is well approximated by
the saturated Comptonization model, its high energy cut-off follows
$kT\approx 2.4$ keV black body. Its independence on the global
mass accretion rate lends support to the
theoretical suggestion by Inogamov\&Sunyaev (1999) that the  boundary layer
is radiation pressure supported. With this assumption we constrain
the gravity on the neutron star surface and its mass and radius.
Equipped with the knowledge of the boundary layer spectrum we attempt
to relate the motion along the Z-track to changes of physically
meaningful parameters. Our results suggest that the contribution of
the boundary layer to the observed emission decreases along the
Z-track from conventional $\sim 50\%$ on the horizontal branch to a
rather small number on the normal branch. This decrease can be
caused, for example, by obscuration of the boundary layer by the
geometrically thick accretion  disk at $\dot{M}\sim\dot{M}_{\rm
Edd}$. Alternatively, this can indicate significant change of the
structure of the accretion flow at $\dot{M}\sim\dot{M}_{\rm Edd}$
and disappearance of the boundary layer as a distinct region of the
significant energy release associated with the neutron star surface.
} {}
   \keywords{accretion, accretion disks --
                instabilities --
                stars:binaries:general --
                stars:neutron --
                X-rays:general  --
                X-rays:binaries
               }
     \maketitle

%
%________________________________________________________________

\section{Introduction}
\label{sec:intro}

Accreting neutron stars in low mass X-ray binaries (LMXB) are
among the most luminous compact X-ray sources in the Galaxy.
At least several of them have luminosities exceeding
$\sim {\rm few}\times 10^{38}$ erg/s and presumably
accrete matter at the level close to the critical Eddington accretion
rate. Early observations of these sources \citep[e.g.][]{toor70}
revealed rather soft  X-ray spectra, indicating that their X-ray
emission is predominantly formed in the optically thick media.
Similar to accreting black holes,  at lower X-ray luminosities (lower
mass accretion rates), $L_{\rm x}\la 5\times 10^{36}$ erg/s,
neutron stars undergo a transition to the hard spectral state
\citep[e.g.][]{barret01}. The energy spectra in this state
point at the low optical depth in the emission region.

In the soft spectral state, the commonly accepted picture of accretion
at not too extreme values of  accretion rate has
two main ingredients -- the  accretion disk (AD) and the boundary
layer (BL).  
While in the disk matter rotates with nearly Keplerian velocities, in
the boundary layer it decelerates down to the spin frequency of the
neutron star and settles onto its surface.
For the typical neutron star spin frequency ($\la500-700$Hz) comparable
amounts of energy are released in  these two regions
\citep{ss86,sibg00}.  
This picture is based on rather obvious qualitative expectations as
well as more sophisticated theoretical considerations and numerical
modeling \citep{ss86,kluzniak,inogamov99,sibg00}.
It has been receiving, however, little direct observational
confirmation.  Due to similarity of the spectra of the accretion disk
and boundary layer the total spectrum has a smooth curved shape, which
is difficult to decompose into separate spectral components
\citep{mitsuda84,white88,disalvo01,done02}.   
This made  application of physically motivated spectral models to the
description of observed spectra of luminous neutron stars difficult,
in spite of very significant increase in the sensitivity of
X-ray instruments. A possible solution was suggested by early results
of \citet{mitsuda84}, which demonstrated the potential of using  
the combined spectral and variability information.

Recently, \cite{gilfanov03} analyzed spectral variability in luminous
LMXBs and showed that in these sources aperiodic and quasi periodic
variability  on  $\sim$ sec -- msec time scales  is caused primarily
by variations of the luminosity of the boundary layer.
Its spectral shape remains nearly constant in the course of
the luminosity variations and is represented by the Fourier frequency
resolved spectrum. Moreover, in the considered  range $\dot{M}\sim
(0.1-1) \dot{M}_{\rm Edd}$ ($\dot{M}_{Edd}$ is the critical Eddington
mass accretion rate) it depends weakly on the global mass accretion
rate and in the limit $\dot{M}\sim \dot{M}_{\rm Edd}$ is close to Wien
spectrum with $kT\sim 2.4$ keV.
Such a behavior is in accord with the predictions of the model by
Inogamov \& Sunyaev (1999), namely, that at sufficiently high
accretion rates, $\dot{M}>0.1  \dot{M}_{Edd}$,  the boundary layer is
radiation pressure dominated and the local radiation flux is close to the
critical Eddington value. Increase of the mass accretion rate
leads to the increase of the emitting area of the BL, while
its vertical structure changes little \citep{inogamov99}.

In this paper we further explore the behavior of the boundary layer
and the accretion disk emission in a number of bright neutron star
LMXBs (both Z and atoll sources) and qualitatively describe their
evolution with change of a mass accretion rate. 
Our study will be based on the results of \cite{gilfanov03}, namely,
that the shape of the boundary layer spectrum is adequately represented
by the frequency resolved spectrum (energy spectrum of the variable
part of the emission) at the Fourier frequencies above $f\ga 1$ Hz.
We will concentrate on
neutron star LMXBs in the soft/high spectral state with luminosities
$L_X\ga 0.5-1 \times 10^{37}$ erg/s. Sources in this state
have very impotant 
property -- their frequency resolved energy spectra do not
depend on Fourier frequency. It is this property which allow us to 
decompose the averaged energy spectrum of the sources into two components.
In hard spectral state the variability is significantly more complicated
\citep[see e.g.][]{freq_res99,mikej00} and we leave it for other studies.

\section{Data}

For our study we used data of RXTE observatory \citep{rxte},
which combines large collecting area with high time resolution.  
We have studied all Z-sources except for GX349+2
which during the majority of RXTE observations was on the
``flaring'' branch of its color-color diagram, supposedly
corresponding to super-Eddington  mass accretion rates
\citep{has_vdk_89}. The behavior of sources at such high  
accretion rates is more complex and is beyond the scope of this
paper. The remaining five Z sources are: Cyg X-2,
GX340+0, Sco X-1, GX 5-1 and GX 17+2.
For these sources we have excluded the time intervals corresponding
to the ``flaring'' branch of their color-color diagram. The atoll
sources were represented in our sample by 4U1608-52 and  4U1820-30 in
their soft/high spectral state, showing high rapid flux
variability.

For the data reduction we used standard programs of package
HEASOFT 5.3 in accordance to the RXTE Guest Observer Facility
recommendations. We have corrected the obtained energy spectra for
the deadtime and the pile up effects ( http://lheawww.gsfc.nasa.gov/
docs/xray/xte/pca/). All subsequent spectral approximations included
interstellar absorption which was fixed at values : \
$N_H=6\times10^{22}$ cm$^{-2}$ for
GX 340+0 \citep{iaria04}, $0.2\times10^{22}$ cm$^{-2}$ for Cyg X-2 \citep{kuulkers97}, 
$0.2\times10^{22}$
cm$^{-2}$ for Sco X-1 \citep{kahn84}, $1.0\times10^{22}$ cm$^{-2}$ for GX 17+2 
\citep{langmeier90},
$3.0\times10^{22}$ cm$^{-2}$ for GX 5-1 \citep{asai94}, $1.0\times10^{22}$ cm$^{-2}$
 for 4U1608-52 \citep{penninx89} and  $0.2\times10^{22}$ cm$^{-2}$ for 4U1820-30 \citep{haberl87}.
The luminosities of all Z sources are approximately $2\times 10^{38}$
erg s$^{-1}$, for 4U1608-52 and 4U1820-30 approximately $10^{37}$ erg
s$^{-1}$.

\begin{figure}
\includegraphics[width=\columnwidth]{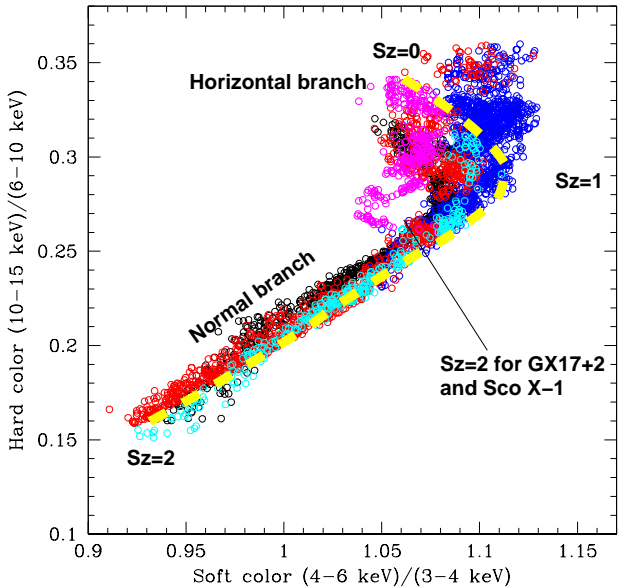}
\caption{Color-color diagram of all Z-sources except for GX349+2.
Only horizontal and normal branch data are shown.
Colors were calculated as ratios of absorption corrected energy fluxes
in the following energy channels 3.0-4.0 keV, 4.0-6.0 keV, 6.0-10.0
keV, 10-15.0 keV.  Dashed line shows the approximation to the
Z-track by an arbitrary smooth curve.
Turning points of the Z-track are marked with $S_z=0,1,2$.
Here and later black points denote GX340+0, red - Cyg X-2, cyan - GX 5-1, magenta - Sco X-1, blue - GX17+2}
\label{ccd}
\end{figure}

The color-color diagrams (CCD) of these sources are presented
in Fig.\ref{ccd}. For construction of the CCDs we have used energy
channels 2.0-4.0 keV, 4.0-6.0 keV, 6.0-10.0 keV and 10.0-15.0 keV.
The hard and soft colors are ratio of energy fluxes of sources
in abovementioned energy bands. Energy fluxes were computed
from spectra averaged over 128 sec intervals.
All fluxes were corrected for absorption.

\begin{figure}
\includegraphics[width=\columnwidth]{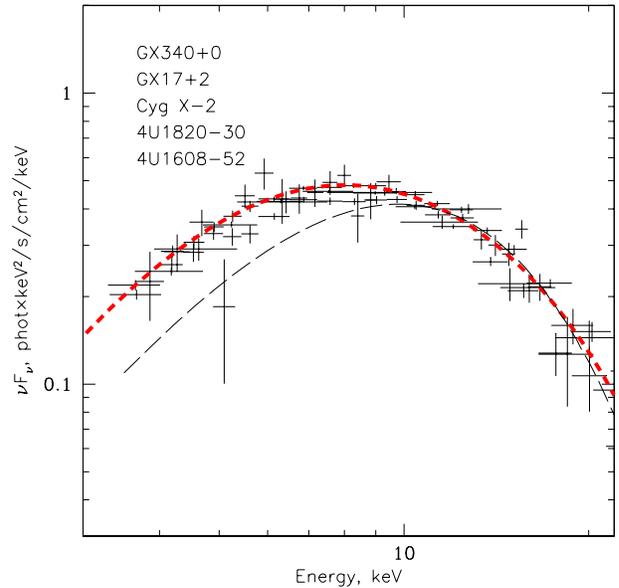}
\caption{Fourier-frequency resolved spectra ($\approx$boundary layer
spectra) of sources from our sample. All spectra were corrected for
the interstellar absorption. The thick dotted line shows the best
fit Comptonization model with $kT_s=1.5$, $kT_e=3.3$ keV, $\tau=5$.
Thin dashed line shows the blackbody spectrum with temperature
$kT_{\rm bb}=2.4$ keV.}
\label{freq_spectra}
\end{figure}

\section{Spectrum of the boundary layer}
\label{sec:bl_spectrum}

The frequency resolved spectra (the energy spectrum of the variable
part of the emission, \citealt{freq_res99})  computed at the frequencies
$f\ga 1-5$ Hz are shown for five sources from our sample in
Fig.\ref{freq_spectra}.  
For Z sources we used only data on the horizontal branch of the CCD where
the amplitude of variability at these frequencies is maximal.
For Sco X-1 and GX 5-1 the RXTE achive does not contain
high time resolution data for the horizontal branch with sufficient
number of energy channels.
The frequency resolved spectrum of 4U1608-52 was taken from
\cite{gilfanov03}.
For plotting purposes, the normalizations of all spectra were adjusted
to match that of GX340+0.

Similarity of the spectra shown in Fig.\ref{freq_spectra} is
remarkable, considering significant difference in the average spectra
and a factor of $\sim 10-20$  spread in the luminosity between atoll
and Z-sources ($\sim 0.1 \dot{M}_{\rm Edd}$ and $\sim \dot{M}_{\rm
Edd}$ correspondingly). This behavior fits in the picture proposed by  
\citet{gilfanov03}. As summarized in the Introduction, they showed, in
particular, that the frequency resolved spectra at  these frequencies
equal the spectrum of the boundary layer emission.

The shape of the frequency resolved ($\approx$ boundary layer)
spectrum can be adequately described by the saturated  
Comptonization. For the sake of comparison with other results
and for convenient parametrization of the  BL spectrum we used the
Comptonization model of Titarchuk (1994). The best fit parameters of
the model fitted in the 3-20 keV range simultaneously to all five
spectra shown in the Fig.\ref{freq_spectra} are:
temperature of seed photons $kT_s=1.5\pm 0.1$, temperature of
electrons $kT_e=3.3\pm0.4$ and the optical depth $\tau=5\pm1$ for slab
geometry. The best fit model is shown by the thick dotted line on
Fig.\ref{freq_spectra}. The temperature of the black body spectrum
describing the shape of the  cutoff in the observed spectrum at
energies $>$13 keV is $kT_{\rm bb}=2.4\pm0.1$ keV (thin dashed line on
Fig.\ref{freq_spectra}).

It would be interesting to follow up on the results of
\cite{gilfanov03} and to consider the evolution of the boundary
spectrum along the Z track from the horizontal to normal branch.
Unfortunately, apart from GX340+0, already considered in
their paper, the variability level on the normal branch of
color-color diagram for other four Z-sources is insufficient to obtain
frequency resolved spectra with reasonable signal-to-noise ratio.

\section{Constrains on NS mass and radius from the BL spectrum}
\label{sec:ns}

The independence of the spectrum of the boundary layer on the
luminosity lends support to the theoretical predictions by
\citet{inogamov99} that the boundary layer is radiation pressure
supported. It means that in the upper layers of the BL 
only the radiation pressure counteracts the gravitational
force. Therefore we can say that every unit area of the BL should
emit Eddington surface flux. At smaller mass accretion rates
($\dot{M}\sim0.1\dot{M}_{\rm Edd}$) the BL occupies only a part of the
NS surface, giving rise to moderate total X-ray luminosities
($L_{\rm x}\sim10^{37}$ erg s$^{-1}$). If mass accretion rate increases
the total BL luminosity approaches the Eddington limit for the NS 
$L_{\rm Edd}\sim2\times10^{38}$ erg s$^{-1}$ $(1.4M_{\odot})^{-1}$.
In this picture, the parameters of the BL emission can be used to
determine the value of the Eddington flux limit on the surface of the
neutron star. As the Eddington flux limit is uniquely determined by
the neutron star surface gravity and the atmospheric chemical
composition, the  neutron star mass and radius can be constrained.

If the boundary layer emitted true black body emission, the radiation
flux of the unit area was determined only by its temperature.
Therefore the observed shape of the BL spectrum, in particular the
best fit black body temperature, could be used to determine the value
of the Eddington flux limit on the neutron star surface. This approach
has been utilized in the context of the Eddington limited
X-ray bursts \citep[e.g.][]{goldman79,marshall82,lewin93,
titarchuk02}. For fully ionized hydrogen atmosphere:
$$
{\sigma T^4\over{c}} {\sigma_{\rm T} \over{ m_{\rm p}}} =
{G M (1-R_{\rm Sch}/R)^{3/2} \over{R^2}}
$$
where $\sigma$  - Stefan-Boltzmann constant, $\sigma_{\rm T}$ -
Thomson cross-section, $T$ -- black body temperature at infinity,
$m_{\rm p}$ -- proton mass, $M$ -- mass and $R$ -- radius of the
neutron star, $R_{\rm Sch}=2GM/c^2$ -- Schwarzschild radius of the
neutron star.
In addition, one would have to take into account that
the value of the Eddington flux is somewhat (by $\sim 10-20$\%)
reduced because of the action of the centrifugal force caused
by the rotation of the boundary layer
\citep{inogamov99}. Rotation of the neutron star at this point is 
not very important unless it is very high (rotational frequency 
$\ga$800-1000 Hz)

\begin{figure}
\includegraphics[width=\columnwidth]{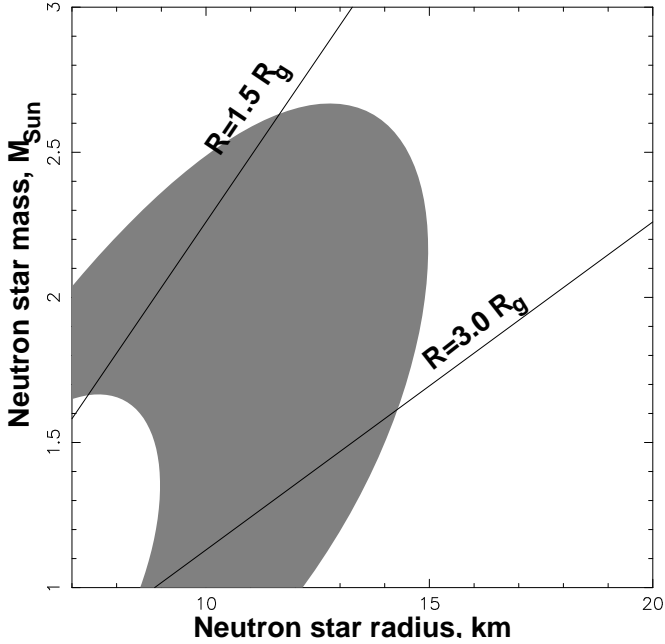}
\caption{Constrains  on the mass and radius of the neutron star in
bright LMXBs. Solid lines denote the regions where neutron star radius $R$
equals to $1.5 R_g$ (radius of photon orbit in Schwarzschild metric)
and $3.0 R_g$ (radius of last stable orbit of an accretion disk in
Schwarzschild metric)}
\label{mr_plot}
\end{figure}

In reality, scatterings are important in the atmosphere of the neutron
star,  therefore, the boundary layer spectrum will differ from the black
body (e.g. \citealt{london86,lewin93}). The radiation transfer problem
in the atmosphere of the neutron star has been intensively
investigated, in particular in the context of X-ray bursts.
Numerical calculations  show that the effects
of scatterings can be approximately accounted for by introducing
the spectral hardening factor which relates the color and the
effective temperatures  of the emission
\citep{ss74,london86,titarchuk94,shimura_takahara95,ross96}.
Typical values of the hardening factor are about $\sim 1.7$.
We used this result in order to make simple estimates of the gravity
on the neutron star surface and to constrain its mass and radius.
In these calculations we assumed the color temperature of the boundary
layer emission $T=2.4 $ keV and considered the range of the
hardening factor values of 1.6-1.8.
We assumed that the centrifugal force reduces
the Eddington flux limit in comparison with non-rotating boundary
layer by 20\%. We also took into account the finite height of the
boundary/spreading layer, about 1 km \citep{inogamov99}. The
calculations were performed for Schwarzschild geometry, hydrogen
atmosphere. The result is shown in Fig.\ref{mr_plot}. The width of the
shaded region is defined by the assumed range of the values of the
hardening factor. Detailed modeling of the emergent spectrum of the
BL will allow to strongly diminish the size of uncertainty region.

Finally we note that the similarity of the spectral shape of the BL
spectrum in different sources (Fig.\ref{freq_spectra}) indicates that
they have close values of the mass and radius, in particular, that
there is no significant difference in the surface gravity between
atoll and Z-sources.  
It also shows that there are no significant differences caused by
variations in the atmospheric chemical abundances between sources.
In particular  we did not find statistically significant difference
between ultracompact compact binary 4U1820-30 and other sources.

\section{Evolution along Z-track}

\begin{figure}
\includegraphics[width=\columnwidth,bb=40 175 570 691,clip]{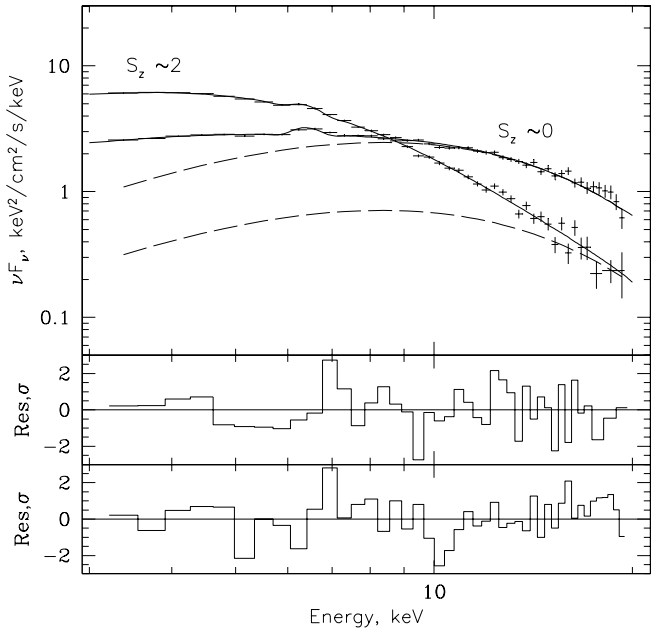}
\caption{Spectra of Cyg X-2 at the beginning of horizontal branch
($S_z\sim0$) and at the end of the normal branch ($S_z\sim2$).
Solid lines are best fit models of the spectra. Dashed lines denote
the boundary layer components}
\label{cygx2_2}
\end{figure}

In order to study the behavior of Z-sources in the color-color
diagram we consider their spectra integrated over 128-sec time
intervals. These spectra are fitted with a simple model consisting
of two components, representing contributions of the boundary layer and
the accretion disk. Also included in the model was the gaussian line
at the energy 6.4 keV. It is not energetically important but
significantly improves quality of the fit. The line parameters will
not be considered here.
The shape of the boundary layer spectrum was approximated by comptt
model with the best fit parameters obtained in the section
\ref{sec:bl_spectrum}. For the accretion disk spectrum we
adopt the multicolor disk blackbody \cite{ss73,mitsuda84}
($diskbb$ model in XSPEC fitting package).

Quality of the spectral fits is uniformly good on the horizontal and
normal branches of the Z-track in the color-color diagram, with the
reduced $\chi^2$  never exceeding  $\sim$1.3 for 41 degrees of
freedom. Relative deviations of the model from the data does not
exceed approximately one percent.
This is  illustrated by Fig.\ref{cygx2_2}, showing two spectra
of Cyg X-2, at the beginning of the horizontal branch  and at the end
of the normal branch.

The model is obviously oversimplified.
Firstly, it relies on the assumption about the shape of the boundary
layer spectrum. This seems to have been established rather well for
the horizontal branch and for the upper part of the normal branch.
For the latter, it has been shown for GX340+0 only, for other
sources the signal-to-noise ratio being insufficient. On the lower
part of the normal branch the variability level and, consequently the
signal-to-noise ratio of the frequency resolved spectra, are too low
to conduct a similar study for any of the sources in our
sample.
Secondly, we chose the diskbb model to represent the spectrum of the
accretion disk. Although the model involves several well known
simplifying assumptions it has been shown to approximately reproduce
the shape of the spectrum of the geometrically thin, optically
thick accretion disk (Merloni, Fabian \& Ross, 2000). The best fit color temperature
obtained in this model can be related with certain accuracy to the
maximum disk temperature.   
Therefore with this model we should be able to separate the disk and
boundary layer components and to estimate  their respective fluxes, as
well as the characteristic temperature in the inner disk.
Keeping in mind the limitations of the spectral model we will attempt
to relate the motion of a source along the Z-track to the changes in
the values of the physically meaningful parameters.
The most important difference of our approach from similar studies
undertaken previously is in a priori knowledge of the shape of the
boundary layer spectrum.

\begin{figure}

\includegraphics[width=\columnwidth]{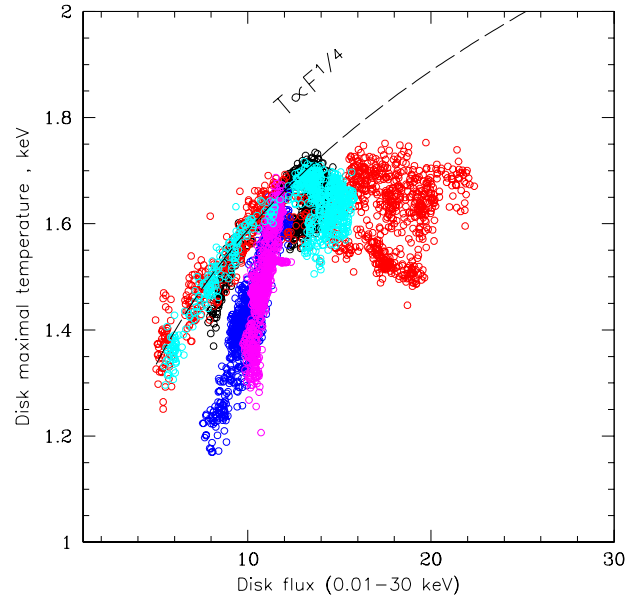}
\caption{Dependence of the disk temperature on its flux
for five Z-sources. The units of flux are $10^{-9}$
erg/s/cm$^2$. Regions corresponding to the horizontal and normal
branches are marked.
Dashed curved  line shows  $T\propto F_{\rm disk}^{1/4}$ dependence.
For GX5-1, GX340+0, GX 17+2 and Sco X-1 the flux values were scaled down
by the factor of 1.3, 1.4, 1.4 and 13 respectively to match Cyg X-2 points
at the end of the horizontal branch}
\label{disk_f_t}
\end{figure}

\subsection{Accretion disk}

The relation between the accretion disk temperature and its flux  
for all five Z-sources is shown in Fig.\ref{disk_f_t} (we emphasize
that plotted along the x-axis is the disk flux, as opposite to the
total flux).
For three of them (GX 5-1, Cyg X-2 and GX 340+0) the disk temperature
on the horizontal branch approximately follows the
$T\propto F_{\rm disk}^{1/4}$ law, which is to be expected for a
standard accretion disk. For Sco X-1 and GX17+2 the temperature
increases with the disk flux faster than the
$T\propto F_{\rm disk}^{1/4}$ law.
For all sources the monotonic relation between the disk temperature
and the disk flux breaks down on the normal branch  and further along
the Z-track the temperature either remains constant or decreases with
the disk flux.  

The significant deviation from $T\propto F_{\rm disk}^{1/4}$ law near
the transition from the horizontal to the normal branch indicates
that the simple accretion disk model in the form represented by
xspec's diskbb becomes inapplicable. The exact reason for this is
not clear. It may happen due to rather small modifications to the
model, e.g. violation of the assumption that the effect of
Comptonization can be represented by the single value of the spectral
hardening factor, constant over the extend of the inner disk as well
as along the Z-track in the color-color diagram. Alternatively, this
might be an  indication of the more significant break down of the
standard geometrically thin,  optically thick picture of the accretion
disk. This would point at the significant modification of the
structure  and geometry of the accretion flow near the transition from
the horizontal to the normal branch, presumably
corresponding to $\dot{M}\sim  \dot{M}_{\rm Edd}$.  

Knowledge of  the boundary layer and the accretion disk
spectra allows one to (crudely) estimate the scale-height of the disk
$H/R$ in the region of the main energy release.
Indeed, for the radiation pressure  supported accretion disk its
height is determined by the equality of the radiation pressure  and
the gravitational force \citep[see e.g.][]{ss73}.
For geometrically thin accretion disk the gravitational force near the
neutron star is $\sim R/H$ times smaller than on the neutron star
surface.
Assuming that the radiation transfer conditions are similar in the
disk and in the boundary layer (i.e. the color hardening
factors are similar, see section \ref{sec:ns}) we find $H/R\sim
(T_{\rm disk}/T_{\rm BL})^4\sim 0.1-0.2$  which is comparable with the
predictions of the simple disk theory for
$\dot{M}\sim\dot{M}_{\rm Edd}$ \citep{ss73,ss76}.  
On the other hand, this estimate can be considered as a crude
consistency check for the obtained values of the boundary layer and
disk temperatures.

\begin{figure}
\includegraphics[width=\columnwidth,bb=40 175 570 691,clip]{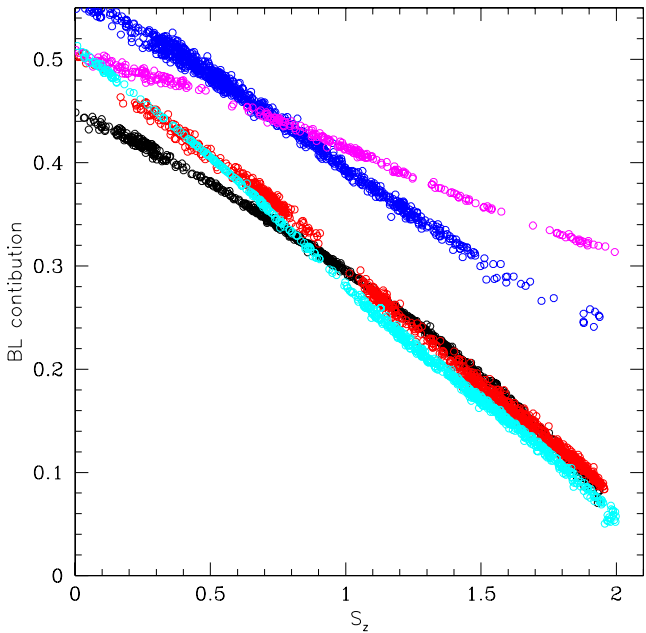}
\caption{Dependence of the boundary layer contribution to the total X-ray
emission of Z sources as a faction of the position on their Z-diagrams}
\label{blcontr}
\end{figure}

\subsection{Boundary layer}

The  dependences of the BL contribution to the total
X-ray emission  on the position on the Z-track are plotted
in Fig.\ref{blcontr}.  The coordinate along the Z-track was defined to
be proportional to the hard color  with the reference points
$S_Z=0,1,2$ defined as shown in Fig.\ref{ccd}.

Statistical uncertainties in the values of the BL fraction
are small and can be neglected, as confirmed by the dispersion of  the
points in Fig.\ref{blcontr}.
More important are the systematic ones associated with the
imprecise knowledge of the shape of the BL spectrum and its possible
variations along the Z-track.  
In order to probe the amplitude of these uncertainties on the
horizontal  branch we explored the range of $1\sigma$ errors of
the parameters of the comptt model obtained in the section 3. The
associated change of the boundary layer fraction
did not exceed $\approx 0.05-0.07$ in the units of
Fig.~\ref{blcontr}.
As for the normal branch, there is some difference between the
frequency resolved spectra of GX340+0 on the horizontal
and normal branch, the latter being better described by the Wien
spectrum with kT=2.4 keV than by the spectrum of saturated
Comptonization (cf. thick and thin dashed lines in
Fig.\ref{freq_spectra}).  
This difference could be related to weak dependence of the
boundary layer spectrum on the global mass accretion rate
\citep{gilfanov03}. Substitution of the comptt model by the
blackbody or Wien  spectrum with $kT=2.4$ keV, decreases the
boundary layer fraction by $\approx 0.1$.
As it is clear from these numbers, systematic uncertainties might
change the Fig.\ref{blcontr} in details but does not affect the
general trend in a significant way.

The Fig.\ref{cygx2_2} and \ref{blcontr} suggest that the boundary
layer fraction decreases along the Z-track and it is smaller on the
normal branch  that on the horizontal branch. As demonstrated above,
this conclusion is rather robust, as long as the assumption regarding
the constancy of the boundary layer spectrum is approximately
correct. Even if the diskbb model is not applicable on the normal
branch any more, the $\sim 5$-fold decrease of the $E\ga15$ keV flux
between $S_z=0$ and $S_z=2$ effectively constrains the  boundary layer
contribution on the normal branch (Fig.\ref{blcontr}).
As the variability at $f\ga 1$ Hz is primarily associated with the
boundary layer emission, the decrease of the boundary layer fraction
along the Z-track also explains well-known decrease of the level of
aperiodic and quasi-periodic variability.

Although no complete physical interpretation of the observed behavior
can be offered, we mention several possibilities. One of these is that
the  general structure of the accretion flow does not change
significantly  and $\sim 50\%$ of the energy is always released on, or
very close to the neutron star surface. The apparent decrease of the
boundary layer  fraction on the normal branch is a result of its
geometrical obscuration by, for example, the geometrically thickened
accretion disk.
An alternative possibility is that at high values of the mass
accretion rate  $\dot{M}\sim\dot{M}_{\rm Edd}$ a significant
modification of the accretion flow structure might occur and its
division into two geometrically distinct parts -- boundary layer and
accretion disk, might become inapplicable.
Namely, due to non-negligible pressure effects the deceleration of the
orbital motion of the accreting  matter from Keplerian frequency to
the neutron star spin frequency would take place in a geometrically
extended region with the  
radial extend of $\Delta R\sim R_{\rm NS}$.
In this case, the  observed decrease of the boundary layer fraction
could reflect actual decrease of the fraction of the energy released
on the neutron star surface with the rest of the energy being released
in the extended transition region.
Finally, it is also possible that the apparent  decrease of the
boundary layer fraction as well as the peculiar behavior of the
accretion disk temperature (Fig.\ref{disk_f_t}) is the consequence of
the complete break down of the model on the normal branch of the
color-color diagram.

Sco X-1 and GX17+2 appear to have higher boundary layer
fraction at the end of the normal branch than other three sources.  
Another indication of the possibly higher contribution of the BL
emission at $S_z\sim 2$ in Sco X-1 and GX17+2 is that they
typically show stronger fast variability at the
normal and flaring branch than Cyg X-2, GX340+0 and GX5-1
\citep[cf. Sco X-1 -like and Cyg X-2 -like sources,][]
{has_vdk_89,kuulkers94,jonker00,piraino02,homan02}.  
The observed difference can be caused, for example, by different
inclination  angles in these systems, as indicated by optical  data
\citep{fomalont01,steeghs02,crampton76,orosz99,kuulkers94}.
It should be mentioned though, that the appearance of the
Fig.\ref{blcontr} is a direct consequence of the definition of the
$S_z$ coordinate, in particular the definition of the $S_z=2$
reference point as the  turning point from the normal to the flaring
branch. If, for example, the BL fraction would be plotted against the
hard color, all five sources will be almost indistinguishable.

\subsection{Shape of Z-diagram}
\label{sec:zdiagram}

Motivated by these results we use the two-component spectral model
introduced in the beginning of this section to explain behavior of
the Z-sources on the color-color diagram.
Unlike before, for the accretion disk spectral component
we adopt here the general relativistic accretion disk model by
Ebisawa et al. (1991) (the {\em grad} model in the XSPEC fitting
package) which explicitly includes dependence on the mass accretion
rate.   
The mass of the compact object and binary system inclination were
fixed at $M_{\rm NS}=1.4 M_{\odot}$ and $i=60^\circ$.  
Using this two-component spectral model we can calculate the position
in the color-color diagram as a function of the mass accretion rate
and the boundary layer fraction.  
This simple model is an attempt to understand  general tendencies of
the Z-diagram, rather than to construct its precise quantitative
description.
In particular, the fact is ignored, that the relation between the
accretion disk temperature and its flux is more complex than predicted
by the standard model of the geometrically thin disk
(Fig.\ref{disk_f_t}).

\begin{figure}
\includegraphics[width=\columnwidth]{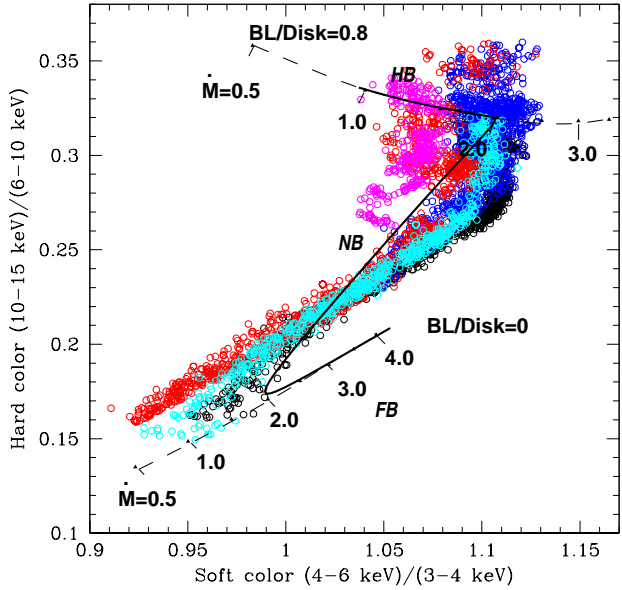}
\caption{The color-color diagram of 5 Z-sources.  Overlayed on the
data are the tracks predicted by the model (section
\ref{sec:zdiagram}). The thick two dashed lines show evolution of the
colors with change of the accretion rate in the disk for two different
values of the BL fraction: 44\% (upper) and zero (lower)
The values of the  mass accretion rates $\dot{M}$ of the disk
component are marked in units of $10^{18}$ g/s.
The thick solid line shows the Z track with transition mass
accretion rate $2\times10^{18}$ g/s.}
\label{z_model}
\end{figure}

The results are presented in Fig.\ref{z_model}. The two dashed
curves in the figure show the evolution of spectral colors with
the increase of the mass accretion rate of the disk component for two
different values of the boundary layer fraction. The upper curve
assumes the BL-to-disk flux ratio of 0.8,  in the lower curve the
BL fraction equals zero.
On each curve positions where the mass accretion rate equals  
$\dot{M}=0.5,1,2,3,4\times 10^{18}$ g/s are marked.
The evolution of colors assuming a rapid change of the BL contribution  
from $F_{\rm BL}/F_{\rm disk}=0.8$ to zero approximately at
$\dot{M}=2\times 10^{18}$ g/s is shown by thick solid line.

The general shape of the Z-track can be reproduced in the model as a
result of variation of two parameters -- the mass accretion rate and
BL fraction.   
The mass accretion rate increases along the Z-track. The Z-shape of
the track is defined by the variation of the BL fraction, which
decreases along the normal branch from the value of $\sim 50\%$
expected in the ``standard'' theories to a small number of the order
of $\sim$zero at the end of the normal branch.

Quantitatively, the exact value of the $\dot{M}$, corresponding to the
transition from the horizontal to the normal branch depends on the
disk model ($grad$) parameters -- the binary system inclination,
the mass of the neutron star and the spectral hardening factor.
For our choice of parameters, it equals $\dot{M}\sim 2\cdot10^{18}$
g/sec, i.e. is of the order of the Eddington critical value for a
$1.4M_\odot$ neutron star, when we can expect that the structure of the
inner accretion flow can start to change. It is interesting to 
note that similar conclusions (decrease of the observable NS 
emission due to possible inflation of the inner disk) were obtained
from studies of simultaneous observations of Z-sources in radio, optical
UV and X-ray ranges \citep[e.g.][]{hasinger90,vrtilek90}

The location of the horizontal branch on the color-color diagram
depends on the boundary layer fraction  at smaller
$\dot{M}$. Theoretically, it could be used to determine the latter
from observations which would be of great help for the
theory. In particular, our spectral model suggests $F_{BL}/F_{AD}\sim
0.8$. In practice, however, the direct comparison of this number with
theoretical predictions for the energy release in the disk and
boundary layer is complicated, because the
observed flux ratio $F_{\rm BL}/F_{\rm disk}$ is modified by
geometrical factors and anisotropy of the AD and BL emission diagrams.

\section{Summary}

Using the archival data of RXTE observations of several bright LMXBs
we study the boundary layer emission and the structure of the
accretion flow in this sources. Our sample includes 2 atoll sources in
the soft spectral state and 5 (all but one) Z-sources.
Their luminosities range from $\sim 10^{37}$ erg/s (atoll sources) to
$\sim 2\cdot 10^{38}$ erg/s (Z-sources).
Our results are summarized below.

\begin{itemize}

\item
We construct Fourier-frequency resolved spectra (energy spectra of
variable emission) at $f\ga 1-5$ Hz for five sources, which data
combines sufficient timing and spectral  resolution.  
Within the statistical uncertainties all the spectra have the same
shape  (Fig.\ref{freq_spectra}).  
This is a  remarkable result as the average spectra of the atoll and
Z-sources differ significantly and their luminosities range from $\sim
10^{37}$ erg/s (atoll sources) to $\sim 2\cdot 10^{38}$ erg/s
(Z-sources).  

This result fits in the picture suggested by  \citet{gilfanov03} and
briefly outlined in the Introduction. This picture implies, in
particular, that the frequency resolved spectrum equals the energy
spectrum of the boundary layer emission. The following is based
on this assumption.

\item
The boundary layer spectrum in atoll sources and on the horizontal
branch of  Z-sources (i.e. at presumably sub-Eddington accretion
rates) can be approximated by the spectrum of saturated Comptonization
(section \ref{sec:bl_spectrum}).
Its high energy ($E>$10 keV) cut-off corresponds to the black body
spectrum with temperature of $kT\sim 2.4$ keV
(Fig.\ref{freq_spectra}).   

\item
Assuming that the boundary layer is radiation pressure supported we
constrain the neutron star mass and radius.
For a $1.4M_{\odot}$ neutron star and the spectral hardening factor of
1.6--1.8 the  NS radii are in the
range of $R_{\rm NS}\sim9-14$ km (Fig.\ref{mr_plot}).

\item
We  attempt to relate the motion of Z-sources along the Z-track to
changes in the values of the physically meaningful parameters.
Our results tentatively suggest that the contribution of the boundary
layer  component to the observed emission decreases along the Z-track
from the conventional value of $\sim 50\%$ on the horizontal branch to
a rather small number at the end of the normal branch.
The main difference of our approach from previous attempts is in the
apriori knowledge of the shape of the boundary layer spectrum. This
allowed us to avoid ambiguity of the spectral decomposition into
boundary layer and disk components which was one of the major problem 
in previous studies.

\end{itemize}

\begin{acknowledgements}
Authors thank Rashid Sunyaev, Nail Inogamov, Eugene
Churazov, Juri Poutanen and Valery Suleimanov for useful discussions.
This research has made use of data obtained through the High Energy
Astrophysics Science Archive Research Center Online Service, provided
by the NASA/Goddard Space Flight Center.
\end{acknowledgements}

\end{document}